\documentclass[11pt,a4paper,english,nofootinbib,,superscriptaddress]{revtex4}
\usepackage{lmodern}

\usepackage[T1]{fontenc}
\usepackage[latin9]{inputenc}
\usepackage{babel}
\usepackage{color}
\usepackage{amsmath}
\usepackage{graphicx}
\usepackage{amssymb}
\usepackage{esint}
\usepackage[unicode=true, pdfusetitle,
 bookmarks=true,bookmarksnumbered=false,bookmarksopen=false,
 breaklinks=false,pdfborder={0 0 1},backref=false,colorlinks=false]
 {hyperref}
\usepackage{amsmath}
\usepackage{amssymb}
\def\b{\begin{equation}} \def\e{\end{equation}}
\def\bd{\begin{displaystyle}} \def\ed{\end{displaystyle}}
\def\ba{\begin{array}} \def\ea{\end{array}}

\def\bee{\begin{enumerate}}
\def\eee{\end{enumerate}}

\def\K{{\cal K}}

\def\1{\mbox{I\hspace{-.15em}1}}

\def\R{{\rm I\hspace{-.15em}R}}

\def\C{\hspace{3pt}{\rm l\hspace{-.47em}C}}

\def\b{\begin{equation}}
\def\e{\end{equation}}
\def\bee{\begin{enumerate}}
\def\eee{\end{enumerate}}

\makeatletter
\usepackage{latexsym}\usepackage{bm}

\makeatother
\begin{document}

\title{Quantum Linear Gravity in de Sitter Universe \\ On Bunch-Davies vacuum state}

\author{M.V. Takook}

\email{takook@razi.ac.ir}

\affiliation{Department of Physics, Razi University,
Kermanshah, Iran}
\affiliation{Department of Physics,
Science and Research branch, \\Islamic Azad University, Tehran,
Iran}

\author{S. Rouhani}
\affiliation{Department of Physics,
Science and Research branch, \\Islamic Azad University, Tehran,
Iran}

\date{\today}

\begin{abstract}

In de Sitter ambient space formalism, the linear gravity can be written in terms of a minimally coupled scalar field and a polarization tensor. In this formalism, the massless minimally coupled scalar field can be quantized on Bunch-Davies vacuum state, that preserves the de Sitter invariance, the analyticity and removes the infrared divergence. The de Sitter quantum linear gravity is then constructed on Bunch-Davies vacuum state, which is also covariant, analytic and free of any infrared divergences. We will conclude that the unique Bunch-Davies vacuum state can be used adequately to constructing the quantum field theory in de Sitter universe.
\end{abstract}

\maketitle
\vspace{0.5cm}
{\it Proposed PACS numbers}: 04.62.+v, 03.70+k, 11.10.Cd, 98.80.H
\vspace{0.5cm}

\section{Introduction} 

One of the problems of quantum field theory in de Sitter space-time is the absence of a unique vacuum state for all massless and massive quantum fields. The linear gravity \cite{taro12} and the massless minimally coupled scalar fields \cite{gareta} are constructed on Gupta-Bleuler vacuum state, which successfully removes the infrared divergence and preserves the de Sitter invariance. For other quantum fields, however, the Bunch-Davies vacuum is adequately the unique vacuum state to constructing the quantum field theory.

The ambient space formalism, $ M_H=\left\{ x\in  \R^5;\;\;\eta_{\alpha \beta}x^\alpha x^\beta=(x^0)^2-\vec x\cdot\vec x-(x^4)^2=-H^{-2}\right\}$, allows us to construct a linear gravity utilizing a polarization tensor and a massless minimally coupled scalar field $\phi_m(x)$ \cite{taro12}:
$$ {\cal K}_{\alpha\beta}(x)={\cal D}_{\alpha\beta}(x,\partial,Z_1,Z_2)\phi_m(x),$$
where $Z_1$ and $Z_2$ are two constant $5$-vectors. They can be fixed in the null curvature limit. These vectors can determine the polarization states. The common method of quantization of scalar field, $\phi_m$, not only breaks the dS symmetry, but also results in appearance of infrared divergence \cite{al,alfo}. However, the Krein space quantization on a Gupta-Bleuler vacuum state removes these unwanted features of scalar field quantization. Although, introducing the negative norm states \cite{gareta} to the theory and thus non-analyticity of the two-point functions is the undesirable effect of this method. It should be noted that intrinsic coordinate system have been used in the Krein space quantization of the scalar field. So, there will be a discordance in extending this approach to linear gravity: polarization tensor ${\cal D}_{\alpha\beta}$ is written in ambient space formalism while the scalar field $\phi_m$ is formulated in intrinsic coordinate system \cite{taro12}.

In the ambient space formalism, however, we can write the massless minimally coupled scalar field $\phi_m$ in terms of  the massless conformally coupled scalar field $\phi_c$ using the following relation \cite{ta14}:
$$  \phi_m(x)= \left[A\cdot\partial^\top + 2 A\cdot x\right]\phi_c(x),$$  
where $A^\alpha$ is a constant five-vector and $\partial^\top_\alpha=\theta_{\alpha\beta}\partial^\beta=\partial_\alpha+H^2 x_\alpha x\cdot \partial$. $\theta_{\alpha \beta}=\eta_{\alpha \beta}+H^2x_{\alpha}x_{
\beta}$ is the transverse projector. The quantum field operator $\phi_c$ is constructed on the Bunch-Davies vacuum state in ambient space formalism \cite{ta14,brgamo,brmo}, so the scalar field $\phi_m$ and the linear gravity ${\cal K}_{\alpha\beta}$ could be constructed on  Bunch-Davies vacuum state. The advantages of the present method relative to the previous ones are: (1) Only one formalism, namely the ambient space formalism, is used for the quantization of various spin fields. (2) There will be a unique vacuum state, {\it i.e.} Bunch-Davies vacuum which is used for the whole quantum field theory. (3) The infrared divergence will not appear in the quantization of the massless minimally coupled scalar field $\phi_m$ and linear quantum gravity ${\cal K}_{\alpha\beta}$. (4) All of the two-point functions are analytic.
 
In the next section the quantization of scalar field $\phi_c$ in ambient space formalism is recalled. The scalar field $\phi_m$ and gravitational field 
${\cal K}_{\alpha\beta}$ are then constructed upon the scalar field $\phi_c$. Finally, conclusion and outlook for further investigation have been presented.

\section{Massless conformally coupled scalar field}

The massless conformally coupled scalar field satisfies the following field equation \cite{brmo,ta97}: 
$$ \left(Q_0^{(1)}-2\right) \phi_c(x)=0, $$
where $Q_0^{(1)}=-H^{-2}\partial^\top\cdot\partial^\top$ is the Casimir operator of the de Sitter group. For simplicity from now on we take $H = 1$ unless circumstances necessitates otherwise. In ambient space formalism, two solutions of the above field equation can be written in terms of dS plane waves $(x\cdot \xi)^{-1}$ and $(x\cdot \xi)^{-2}$, where $\xi^\alpha$ lies in the positive cone $C^+=\left\{\xi^\alpha\in R^5|\; \xi \cdot\xi=0, \; \xi^0>0 \right\}$ \cite{brgamo,brmo}. These solutions are not well defined globally in de Sitter space-time. In order to obtain well defined solutions we must define them in the complex de Sitter space-time. The complex de Sitter space-time is defined by \cite{brgamo,brmo}:
$$ M_H^{(c)}=\left\{ z=x+iy\in  \C^5;\;\;\eta_{\alpha \beta}z^\alpha z^\beta=(z^0)^2-\vec z\cdot\vec z-(z^4)^2=-H^{-2}\right\}$$
\b =\left\{ (x,y)\in  \R^5\times  \R^5;\;\; x^2-y^2=-H^{-2},\; x\cdot y=0\right\}.\e
So, the field operator can be well defined in the complex dS space-time by using the analytic complex de Sitter plane waves \cite{ta14}:
\b \phi_c(z)=\sqrt{ c_0 }\int_{S^3}   d\mu({\bf \xi}) \left\lbrace\; a({\bf\tilde{\xi}})(z\cdot\xi)^{-2}
         +a^\dag({\bf \xi})(z\cdot\xi)^{-1} \right\rbrace ,\e
where $\xi^\alpha=(1, \vec \xi, \xi^4)$, $\tilde \xi^\alpha=(1, -\vec \xi, \xi^4)$. The vacuum state is defined as \cite{ta14}:
$$ a({\bf \xi})|\Omega>=0,\;\; a^\dag({\bf \xi})|\Omega>=|\xi>, \;\;< \xi' |\xi>=\delta_{S^3}(\xi-\xi'), \;\;\int_{S^3}   d\mu({\bf \xi})\delta_{S^3}(\xi-\xi') =1.$$ 
The vacuum state $|\Omega>$ in this case is exactly equivalent to the Bunch-Davies vacuum state \cite{brgamo,brmo}. The notations are defined explicitly in \cite{ta14}. 

The analytic two-point function in terms of complex de Sitter plane waves is \cite{brgamo,brmo}:
\b  \label{tpfscinint}  W_c(z,z')=\left<\Omega|\phi(z)\phi(z')|\Omega\right>=c_0\int_{S^3}d\mu(\xi) (z\cdot\xi)^{-2}(z'\cdot\xi)^{-1},\e 
and $c_0$ will obtain using the local Hadamard condition. One can easily calculate (\ref{tpfscinint}) in terms of the generalized Legendre function \cite{brmo}:
\b \label{atpfc} W_c(z,z')=\frac{-iH^2}{2^4\pi^2} P_{-1}^{(5)}(H^2 z\cdot z')=
\frac{H^2}{8\pi^2}\frac{-1}{1-{\cal Z}(z,z')}=\frac{H^2}{4\pi^2}(z-z')^{-2}, \e
where ${\cal Z}(z,z')=-H^2 z\cdot z'$. The Wightman two-point function ${\cal W}_c(x,x')$ is the boundary value (in the sense of its interpretation as a distribution function, according to the theorem A.2 in \cite{brmo}) of the function $W_c(z, z')$ which is analytic in the ``tuboid'' ${\cal T}_{12}$ of $M_H^{(c)}\times M_H^{(c)}$ \cite{brmo}. The tuboid above $M_H^{(c)}\times M_H^{(c)}$ is defined by
\b \label{tuboid} {\cal T}_{12}=\left\{ (z,z');\;\; z\in {\cal T}^+,z' \in {\cal T}^- \right\}, \e
where ${\cal T}^\pm$ are called forward and backward tubes of the complex dS space $X_H^{(c)}$  \b {\cal T}^\pm=T^\pm\cap M_H^{(c)}.\e
$T^\pm= \R^5+iV^\pm$ are the forward and backward tubes in $ \C^5$. The domains $V^\pm$ stem from the causal structure on $M_H$:
\b \label{v+-} V^\pm=\left\{ x\in \R^5;\;\; x^0\stackrel{>}{<} \sqrt {\parallel \vec x\parallel^2+(x^4)^2}\right\}.\e 
For more details, see \cite{brmo}. The boundary value is defined for
$z =x+iy\in {\cal T}^-$ and
$z'=x'+iy'\in {\cal T}^+$ by
   $$ {\cal Z}(z,z')={\cal Z}(x,x')-i\tau\epsilon(x^0,x'^0),$$
where $y=(-\tau,0,0,0,0)\in V^-$, $y'=(\tau,0,0,0,0)\in V^+$ and $\tau
\rightarrow 0$. Then, one obtains \cite{brmo,ta97,chta}:
  $$ {\cal W}_c(x,x')=\frac{-H^2}{8\pi^2}\lim_{\tau \rightarrow 0}\frac{1}{1-{\cal
Z}(x,x')+i\tau\epsilon(x^0,x'^0)}$$ \b \label{stpci2}
=\frac{-H^2}{8\pi^2}\left[
P\frac{1}{1-{\cal Z}(x,x')}
 -i\pi\epsilon(x^0,x'^0)\delta(1-{\cal Z}(x,x'))\right],\e
where $P$ was used for principal part. ${\cal Z}(x,x')$ is geodesic distance between two points $x$ and $x'$ on the de Sitter hyperboloid:
$${\cal Z}(x,x')=-H^2 x\cdot x'=1+\frac{H^2}{2} (x-x')^2,  $$
and   \b \epsilon (x^0-x'^0)=\left\{\begin{array}{clcr} 1&x^0>x'^0
 \\
  0&x^0=x'^0\\  -1&x^0<x'^0\\    \end{array} \right. .\e

\section{Massless minimally coupled scalar field}

The massless minimally coupled scalar field equation is:
$$ Q_0^{(1)} \phi_m(x)=0.$$ 
This field equation is invariant under the transformation 
$$ \phi'_m(x)=\phi_m(x)+ \mbox{const.}\;.$$ 
The solutions of the field equation in terms of the de Sitter plane waves are $(x\cdot\xi)^{-3}$ and $ (x\cdot\xi)^{0}$. The constant solution $((x\cdot\xi)^0=$constant), causes the zero mode problem \cite{alfo,gareta}. With just one solution $((x\cdot\xi)^{-3} )$, it is not possible to establish a proper covariant quantum field operator on the Hilbert space constructed on a unitary irreducible representation of the dS group   \cite{al,gareta,alfo}. 
Nevertheless, one can associate a massless minimally coupled scalar field with an indecomposable representation of the dS group \cite{gareta}. Using the following identities 
$$Q_0^{(1)} \partial^\top_\alpha\phi(x)- \partial^\top_\alpha Q_0^{(1)} \phi(x)=2\partial^\top_\alpha \phi(x)+2x_\alpha Q_0^{(1)}\phi(x),$$
$$Q_0^{(1)} x_\alpha\phi(x)- x_\alpha Q_0^{(1)}\phi(x) =-2\partial^\top_\alpha\phi(x)-4x_\alpha\phi(x),$$
with $\phi$ as an arbitrary scalar field, one can prove the existence of a magic relation between the minimally coupled and the conformally coupled scalar fields in the dS ambient space formalism \cite{ta14}
\b \label{msfincsf}  \phi_m(x)= \left[A\cdot\partial^\top + 2 A\cdot x\right]\phi_c(x). \e  
 $A^\alpha$ is a constant five-vector, which is determined by a representation of the dS group. Such representations can be constructed as the product of two representations of the dS group: $(1)$ the scalar complementary representation related to the conformally coupled scalar field \cite{ta14}, and $(2)$ a five-dimensional trivial representation with respect to $A_\alpha^{(l)}$ \cite{gaha}. For a thorough investigation regarding the five existing polarization states $l = 0,1, 2, 3, 4$, the reader may refer to \cite{gaha}. This subject will not be pursued here since the quantum field operator can be constructed from the confrormally coupled scalar field and the identity (\ref{msfincsf}).

Apart from the polarization five-vector $A_\alpha^{(l)}$, the quantum field operator in complex de Sitter space-time can be defined properly from the quantum field operator of conformally coupled scalar field:
$$ \phi_m(z)=\sqrt{ c_0 } \int_{S^3}   d\mu({\xi})\sum_{l=0}^4\left[A^{(l)}\cdot\partial^\top + 2 A^{(l)}\cdot z\right]  \left\lbrace\; a(
{\bf \tilde{\xi}})(z\cdot\xi)^{-2}
       +a^{\dag}({\xi})(z\cdot\xi)^{-1}
        \; \right\rbrace $$
    $$    =\sqrt{ c_0 } \sum_{l=0}^4 \int_{S^3}   d\mu({\xi}) \left\lbrace\; a(
{\bf \tilde{\xi}})\left[-2(A^{(l)}\cdot\xi^\top)(z\cdot\xi)^{-3} + 2 (A^{(l)}\cdot z)(z\cdot\xi)^{-2}\right] \right.$$ \b \label{mcsfico}
      \left. +a^{\dag}(
{\xi})\left[-(A^{(l)}\cdot\xi^\top)(z\cdot\xi)^{-2} + 2 (A^{(l)}\cdot z)(z\cdot\xi)^{-1}\right]
        \; \right\rbrace\equiv \Phi_m(z,A).\e 
  
The analytic two-point function can be defined on the vacuum state of the conformally coupled scalar field or Bunch-Davies vacuum state as:
$$ W_{m}^H(z,z') =<\Omega| \phi_m(z)\phi_m(z')|\Omega>$$ $$= \sum_{l=0}^4 \sum_{l'=0}^4 \left[A^{(l)}\cdot\partial^\top + 2 A^{(l)}\cdot z\right]\left[A^{(l')}\cdot\partial'^\top + 2 A^{(l')}\cdot z'\right]{\cal W}_c(z,z').$$
The explicit form of this function depends on the chosen representation. As a simple case, one can choose \cite{ta14}:
\b \label{zpolar}  \sum_{l=0}^4 \sum_{l'=0}^4 A^{(l)}_\alpha A^{(l')}_\beta=\eta_{\alpha\beta},\;\;\; A^{(l)}\cdot A^{(l')}=\eta ^{ll'},\e
which results to the constant trivial solution:
\b  W_{m}(z,z')= \left[\partial^\top\cdot\partial'^\top +2z\cdot\partial'^\top + 2 z' \cdot \partial^\top+ 4 z\cdot z'\right] W_c(z,z')=\mbox{constant},\e
with $W_c$ being the analytic two-point function of conformally coupled scalar field (\ref{atpfc}). By using the following identities $${\cal Z}=-H^2z\cdot z',\;\;\;\frac{\partial}{\partial z^\alpha}=-H^2z'_\alpha\frac{d}{d{\cal Z}}, \;\;\;\; \partial^\top_\alpha=\left(z_\alpha {\cal Z}-z'
_\alpha\right)\frac{d}{d{\cal Z}},$$ 
$$ \partial^\top\cdot\partial'^\top=\left(-3+{\cal Z}^2\right)\frac{d}{d {\cal Z}}+{\cal Z}\left(1-{\cal Z}^2\right)\frac{d^2}{d {\cal Z}^2},\;\; z'\cdot\partial^\top=\left(1-{\cal Z}^2\right)\frac{d}{d{\cal Z}},$$
the above relations can be proved. The general two-point function can also be written as:
\b\label{atpfm}  W_{m}(z,z')=  \left[A\cdot\partial^\top + 2 A\cdot x\right] \left[A\cdot\partial'^\top + 2 A\cdot x'\right] W_c(z,z').\e
The general two-point function in the real de Sitter space is the boundary value of the analytic two-point function $ W_{m}(z,z')$ (\ref{atpfm}):
$$ {\cal W}_{m}(x,x') =\frac{-H^2}{8\pi^2} \left[A\cdot\partial^\top + 2 A\cdot x\right] \left[A\cdot\partial'^\top + 2 A\cdot x'\right]$$ \b \label{mth} \times \left[ P\frac{1}{1-{\cal
        Z}(x,x')}-i \pi\epsilon (x^0-x'^0)\delta(1-{\cal Z}(x,x'))\right].\e
In conclusion, the analytic two-point function (\ref{mth}) is free of any infrared divergences.    

\setcounter{equation}{0}
\section{Linear quantum gravity }

In a previous paper, we showed that in  the de Sitter ambient space formalism, the linear gravity can be written in terms of the massless minimally coupled
scalar field $\phi_m$ (for gauge fixing parameter $c=\frac{2}{5}$) \cite{taro12}: \b \K_{\alpha \beta}(x)={\cal
D}_{\alpha \beta}(x,\partial,Z_1,Z_2)\phi_m(x,A),\e where
$$ {\cal D}(x,\partial,Z_1,Z_2)=\left(-\frac{2}{3}\theta Z_1\cdot+{\cal
S} Z_1^\top+\frac{1}{9 }D_2 \left[H^2 xZ_1\cdot-Z_1\cdot \partial^\top
+\frac{2}{3}H^2 D_1 Z_1\cdot\right]\right)$$ \b\left(  Z_{2}^\top-\frac{1}{2}
D_{1}\left[Z_2\cdot
\partial^\top+2H^2x\cdot Z_2\right]\right).\e
The operator $D_1$ is $D_1=H^{-2} \partial^\top$ and the operator $D_2$ is the generalized gradient \b
D_2K=H^{-2}{\cal S}(\bar
\partial-H^2x)K,\e
where ${\cal S}$ is the symmetrizer operator and $K$ is a vector field.  $Z_1$ and  $Z_2$ are the constant five-vectors. They determine (or in other words ''fix'') the specific representation of the de Sitter group. The linear gravitational field operator in complex de Sitter space-time is define by:
$$  \K_{\alpha \beta}(z)=\sqrt{ c_0 } \int_{S^3} d\mu({\xi}) {\cal
D}_{\alpha \beta}(z,\partial,Z_1,Z_2) \left[A\cdot\partial^\top + 2 A\cdot z\right] \left[a(
{\bf \tilde{\xi}})(z\cdot\xi)^{-2}
       +a^{\dag}({\xi})(z\cdot\xi)^{-1}
       \right]. $$
The bi-tensor two-point function can be written in the following form  $(c=\frac{2}{5})$ \cite{taro12} \b {\cal
W}_{\alpha\beta \alpha'\beta'}(x,x')=\Delta_{\alpha\beta
\alpha'\beta'} (x,x'){\cal W}_{m}(x,x'), \e
where ${\cal W}_{m}$ is the two-point function for the massless
minimally coupled scalar field (\ref{mth}) and 
$$ \Delta(x,x')=-\frac{2}{3}{\cal S'}\theta
\theta'\cdot\left(\theta\cdot\theta'
    -{\frac{1}{2}}D_{1}\left[2H^2 x\cdot\theta'+\theta'\cdot\partial^\top\right]\right)$$
$$ +{\cal S}{\cal S}'\theta\cdot\theta'\left(\theta\cdot\theta'
    -{\frac{1}{2}}D_{1}\left[2H^2 x\cdot\theta'+\theta'\cdot\partial^\top\right]\right)$$ \b +\frac{H^2}{9} {\cal
S}'D_2 \left(\frac{2}{3}D_1\theta'\cdot + x\theta'\cdot -
H^{-2}{\theta'}\cdot{\partial^\top}\right)\left(\theta\cdot\theta'
    -{\frac{1}{2}}D_{1}\left[2H^2 x\cdot\theta'+\theta'\cdot\partial^\top\right]\right).\e 
This two-point function which defined completely on the de Sitter ambient space formalism, is analytic, de Sitter covariant and free of any infrared divergences. It is constructed on Bunch-Davies vacuum state.

\section{Conclusion and outlook}

In a series of papers, we constructed the massless and massive fields with spin$=0,\frac{1}{2},1, \frac{3}{2},2$ in de Sitter ambient space formalism (for review see \cite{ta14}). It is shown that there is a unique Bunch-Davies vacuum state in this space which upon the quantization of the massless minimally coupled scalar field and the linear quantum gravity can be achieved. The infrared divergences had been removed from the quantization of scalar field $\phi_m$ and  linear quantum gravity ${\cal K}_{\alpha\beta}$. The two-point functions are all analytic in this construction. As the result, in de Sitter ambient space formalism the quantum field theory is completely unitary and analytic then a unitary supergravity in this formalism seems quite plausible, which will be studied in a forthcoming paper.

\vspace{0.5cm} \noindent {\bf{Acknowlegements}}: We would like to thank Edouard Br\'ezin, Jean-Pierre Gazeau, Eric Huguet, Jean Iliopoulos, Richard Kerner and Jacques Renaud for their helpful discussions.

\end{document}